\documentclass[aasms4,usenatbib]{mn2e}
\usepackage{epsfig}
\usepackage{graphicx}
\usepackage{amssymb}
\begin{document} 

\title[Population synthesis studies of NSs with magnetic field decay]
{Population synthesis studies of isolated neutron stars with  
magnetic field decay}

\author[S.B. Popov, J.A. Pons, J.A. Miralles, P.A. Boldin, B. Posselt]
{ S.B. Popov $^{1}$, J.A. Pons $^{2}$, J.A. Miralles $^{2}$,   P.A. Boldin $^{3}$, B. Posselt $^{4}$
\thanks{E-mail: jose.pons@ua.es(JAP); polar@sai.msu.ru (SBP)}\\
$^1${\sl Sternberg Astronomical Institute, Universitetski pr. 13,   
Moscow, 119991, Russia} \\
$^2$ {\sl Departament de F\'{\i}sica Aplicada, Universitat d'Alacant, Ap. Correus 99, 03080 Alacant, Spain}\\
$^3${\sl Moscow Engineering Physics Institute (state University), Moscow,
Russia}\\
$^4${\sl Harvard-Smithsonian Center for Astrophysics, 60 Garden Street, MS
67 Cambridge, MA 02138 USA}}
\date{Accepted ......  Received ......; in original form ......
      %(MNRAS, 2009, in press)
      }

\maketitle

\begin{abstract}
We perform population synthesis studies of different types of neutron stars
(thermally emitting isolated neutron stars, normal radio pulsars, magnetars)
taking into account the magnetic field decay and using results from the most
recent advances in neutron star cooling theory. For the first time, we
confront our results with observations using {\it simultaneously} the Log N
-- Log S distribution for nearby isolated neutron stars, the Log N -- Log L
distribution for magnetars, and the distribution of radio pulsars in the $P$
-- $\dot P$ diagram. For this purpose, we fix a baseline neutron star model
(all microphysics input), and other relevant parameters to standard values
(velocity distribution, mass spectrum, birth rates ...), allowing to vary
the initial magnetic field strength. We find that our theoretical model is
consistent with all sets of data if the initial magnetic field distribution
function follows a log-normal law with $<\log (B_0/[G])>\sim 13.25$ and
$\sigma_{\log B_0}\sim 0.6$. The typical scenario includes about 10\% of
neutron stars born as magnetars, significant magnetic field decay during the
first million years of a NS life (only about a factor of 2 for low field
neutron stars but more than an order of magnitude for magnetars), and a mass
distribution function dominated by low mass objects. This model explains
satisfactorily all known populations. Evolutionary links between different
subclasses may exist, although robust conclusions are not yet possible.
\end{abstract}


\section{Introduction}

 At the present moment our knowledge about neutron star (NS) evolution is
 an intriguing puzzle. We know many observational manifestations of young
 isolated NSs: radio pulsars (PSRs); central compact objects in supernova
 remnants (CCOs in SNRs);  rotating radio transients (RRATs); radio-quiet thermally emitting
 isolated NSs, also known as X-ray dim isolated NSs (XDINS) or the Magnificent
 Seven (M7), 
and the observational manifestations of magnetars --soft gamma-ray repeaters (SGRs) and anomalous X-ray
 pulsars (AXPs).
 Reasons for this apparent diversity as well as possible links between
 different classes are not entirely clear 
 (see e.g. \cite{psr2005,WT2006,Kaspi2007,Haberl2007,Zane2007,rrat2008} 
for recent reviews about the different subclasses).

In the past few years we have learned that some NSs can show different types
of activity, transiting from class to class. For example, PSR J1846-0258, known for some time
as a normal pulsar,
demonstrated outbursts typical for AXPs or/and SGRs \citep{kumar2008,
gavriil2008}. 
In addition, the total energy release by the object became larger than the rotational energy losses, 
which according to the original classification discussed in
\cite{TD1996} should place it in the AXP list.
Thus, for the first time, we have an example of transformation of a PSR into a magnetar.  
Several of the SGRs do not show any bursting activity for many years, and if we had not enough 
information about their violent past, we
would have classified them as AXPs.
Some of RRATs were shown to emit normal radio
pulsar emission \citep{deneva2009,rrats2009}. The transient AXP XTE J1810-197 and AXP
1E 1547.0-5408 demonstrated radio pulses
\citep{camilo2006, camilo2007}. One of the RRATs J1819-1458 shows thermal
properties much similar to the M7 \citep{reynolds2006}. Hence, divisions between
some subpopulations of young isolated NSs (or at least some of their
representatives) can be illusive. On the other hand, young, low-field CCOs
\citep{halpern2007}, normal PSRs ($B\sim 10^{12}$~G, i.e. Crab and Vela-like), 
and SGRs clearly represent NSs born with different properties.

Our brief observational record of NSs ($\sim 40$ years at most), low
statistics in many cases, and selection effects do not allow to draw a
coherent picture of NS ``sociology'' just from observations. From the theoretical point of view,
our understanding of the SN explosion mechanism is not precise enough to provide
a solid model of initial parameters of NSs  and evolutionary models
(thermal, magnetic field, and spin evolution) are related to extremely
complicated physical problems (superfluidity and superconductivity in dense
matter, electrodynamics in superstrong magnetic fields, etc.) which usually
leads to inconclusive results.

In our opinion, it is necessary not only to compare observations with models to verify individual objects,
but also to confront theoretical calculations with observational data
via population synthesis techniques taking into account as many classes of
NSs as possible. Joint constraints by means of simultaneous comparison of 
theoretical models with different subpopulations should be
derived to form a {\it population mosaic}. Numerous degrees of freedom in
modern evolutionary models must be compensated by different observational
tests. Several important studies in this area appeared in recent years,
see for example \cite{fgkaspi2006, keane2008}, and references therein. In this paper we
continue to follow this lane but with an important difference: we attempt to constrain 
our model and check its consistency using at the same time different populations. This is, to our knowledge, the
first time that a multilateral approach is employed. Previous works that focused in a specific
NS observational class, ignored that the parameters obtained may be in contradiction with
the properties of a different class of objects. For examples, models without magnetic field 
decay are clearly in contradiction with the existence of magnetars. We try to give a step
forward in the direction towards a NS unified model by using at the same time different
populations.

Among the different physical ingredients needed to properly model the
thermal evolution of NSs, we emphasize that heat transport in the NS crust
plays a crucial role during the first million years of its thermal evolution
\citep{Yakovlev2004}, which is typically the period during which NSs are
detectable with current  X-ray instruments. Multi-wavelength observations in the
soft X-ray, UV, and optical bands of the thermal emission from a NS's
surface now provide a real opportunity to probe the internal physics of NSs
(see \citealt{Page2006} for a general overview).

Remarkably, all isolated nearby compact X-ray sources that have been
detected also in the optical band (RXJ 185635-3754, RX J0720.4-3125, RX
J1308.6+2127, and RX J1605.3+3249) have a significant optical excess
relative to the extrapolated X-ray blackbody emission \citep{Haberl2007}. 
An  inhomogeneous surface temperature distribution can accommodate this optical
excess and can arise naturally if heat conduction in the NS crust is
anisotropic due to the presence of a large magnetic field
\citep{Geppert2004,Geppert2006,Azorin2006a}.
Alternative models are based on anisotropic radiation from magnetized
atmospheres, as in \citet{Ho2007}, or condensed surfaces as in \citet{Azorin2005}. 
None of the explanations (only temperature anisotropy vs. physical origin) are
entirely satisfactory and probably a combination of both effects is needed.
Heat conduction can also
influence other observable aspects of accreting NSs in low mass X-ray binaries,
including their quiescent luminosity and the superburst recurrence
time-scales \citep{Brown1998}. For this reason, one of the goals of this
paper is to revisit former population synthesis studies using new NS
evolution models that include anisotropic heat transport in NSs crusts and
magnetic field evolution \citep{Aguilera2008b,Aguilera2008a,PMG2009}.

In this paper we want to study if the {\it patchy} view of
young NSs subpopulations can be explained by a unique set of
smooth distributions of the most important parameters, among which the 
the magnetic field distribution plays the main role.\footnote{Note for the
arXiv version: After the
final (second revision) version of this paper has been submitted, an
important e-print by Kaplan and van Kerkwijk appeared (arXiv: 0909.5218).
In this article the authors give new observational arguments in favor of
the large role of the field decay in the
evolution of the Magnificent Seven. Many of
our conclusions coincide with those given by Kaplan and van Kerkwijk.}
Our main idea here is to  use several different tests to
confront theoretical predictions and observations. For this purpose, the
paper is organized as follows. In the next section we briefly describe the
model of magneto-thermal evolution of NSs that we use. 
With this model, one has the cooling behavior and magnetic field (and therefore
period) evolution of NS.
In Sec.~3, we describe the population synthesis technique used for Log N~--~Log S
calculations of close-by cooling NSs which thermal emission has been
detected and which temperature has been estimated. With this, we can partially constrain the
initial magnetic field distribution. Then, in Sec.~4, we discuss
the Log N~--~Log L distribution of the population of magnetars in the
Galaxy. We show that the model constrained in the previous section is
also consistent with the flux distributions of the extrapolated magnetar population.
To finish the presentation of our results, in Sec.~5 we use the
population of radio pulsars in the $P-\dot P$ diagram to put additional
constraints on the properties of NSs. Explicitly, the current period and magnetic field
distributions of rotation-powered pulsars constrain the NS initial magnetic field distribution 
and break the degeneracy in the parameter space obtained in previous section. 
 Sec.~6 is devoted to the final remarks  and to discuss uncertainties of the models we use and  future
prospects.

%%%%%%%%%%%%%%%%%%%%%%%%%%%%%%%%%%%%%%%%%%
\section{Magnetic field decay and cooling model}

Very often thermal and magneto-rotational evolution of NSs are treated
separately. However, in the case of young ($< 1$ Myr) NSs with magnetic fields $>10^{13}$ G
this is incorrect, because temperature affects the electrical resistivity,
and therefore the magnetic field evolution, while field decay provides an additional energy
source that modifies the temperature of the star.
Although for the average radio pulsar population (old and relatively low field)
this effect is probably not very important \citep{fgkaspi2006}, there is some
observational evidence of the interplay of the magnetic field
and temperature during early stages of NS evolution. As discussed in \cite{PonsLink2007}, there is a
strong correlation between the inferred magnetic field and the surface
temperature in a wide range of magnetic fields: from magnetars ($B \geq
10^{14}$ G), through radio-quiet isolated NSs ($B \simeq 10^{13}$ G) down to
some ordinary PSRs ($B \leq 10^{13}$ G). The main conclusion was that,
rather independently from the stellar structure and the matter composition,
the correlation can be explained by heating from dissipation of currents in
the crust on a timescale of $\simeq 10^{6}$ yrs.

This observed correlation has been confirmed later by more detailed 2D
cooling simulations combining the insulating effect of strong non radial
fields with the additional source of heating due the Ohmic dissipation of
the magnetic field in the crustal region
\citep{Aguilera2008b,Aguilera2008a,PMG2009}. It was shown that, during the
neutrino cooling era and the early stages of the photon cooling era, the
feedback between Joule heating and magnetic diffusion is strong, resulting
in a faster dissipation of the stronger fields. As a consequence, all
neutron stars born with fields over a critical value ($>5 \times 10^{13}$ G)
reach similar field strengths ($\approx 2-3 \times 10^{13}$ G) at late
times. Irrespective of the initial magnetic field strength, the temperature
becomes so low after a few million years that the magnetic diffusion
timescale becomes longer than the typical ages of PSRs, thus
apparently resulting in no dissipation of the magnetic field in old NS.
Another interesting result was that the effective temperature of models with
strong internal toroidal components is systematically higher than that of
models with purely poloidal fields, due to the additional energy reservoir
stored in the toroidal field that is gradually released as the field
dissipates.

In this paper, we have employed cooling curves obtained from an updated
version of the 2D cooling code described in \cite{PMG2009}, that also includes
the effect of superfluid heat conduction in the inner crust
\citep{Aguilera2009}. We have used a Skyrme-type equation of state (EoS) at
zero temperature describing both, the NS crust and the liquid core, based on
the effective nuclear interaction SLy \citep{Douchin2001}. It is a purely hadronic,
relatively stiff EOS that gives typical proper radii in the range $\approx 11.3-11.8$ km
(the radius observed at infinity would be 10--30 \% larger depending on the mass), 
while the crust thickness varies from 0.6 to 1.2 km also depending on the mass of the NS.

We refer to section 4 in \cite{Aguilera2008b} for more details about the cooling models
(neutrino emissivities, equation of state, thermal conductivities, etc.).
Details about the magnetic field initial geometry can be found in
 \cite{PMG2009}. 
Our baseline initial model consists of a crustal--confined magnetic field with a
poloidal component, parameterized by the value of the radial component at the magnetic pole ($B$)
combined with a  toroidal component with a maximum value of twice $B$  
(see Eqs. (11) and (13) of \cite{Aguilera2008b}). We find that, although
the amplitudes of both fields  are of the same
order of magnitude, the contribution of the toroidal field to the total
magnetic energy is $\lesssim10\%$, because this field is non
vanishing only in a finite region of the star. This model is in agreement with the results
obtained in recent studies of magnetic equilibrium configurations \citep{mhd1,mhd2}.

%%%%%%%%%%%%%%%%%%%%%%%%%%%%%%%%%%%%%
\begin{figure*}
\includegraphics[width=230pt,angle=0]{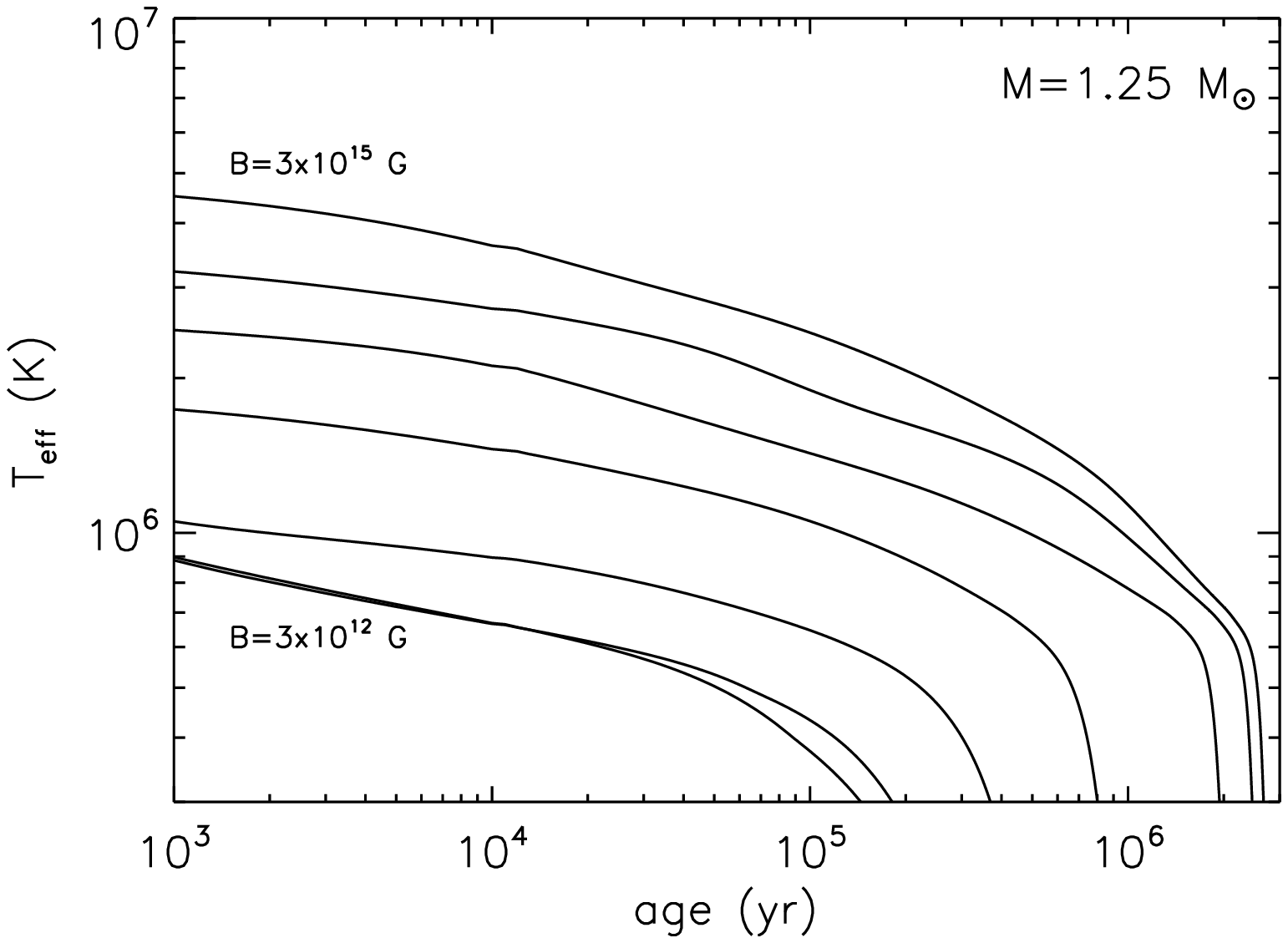} 
\includegraphics[width=230pt,angle=0]{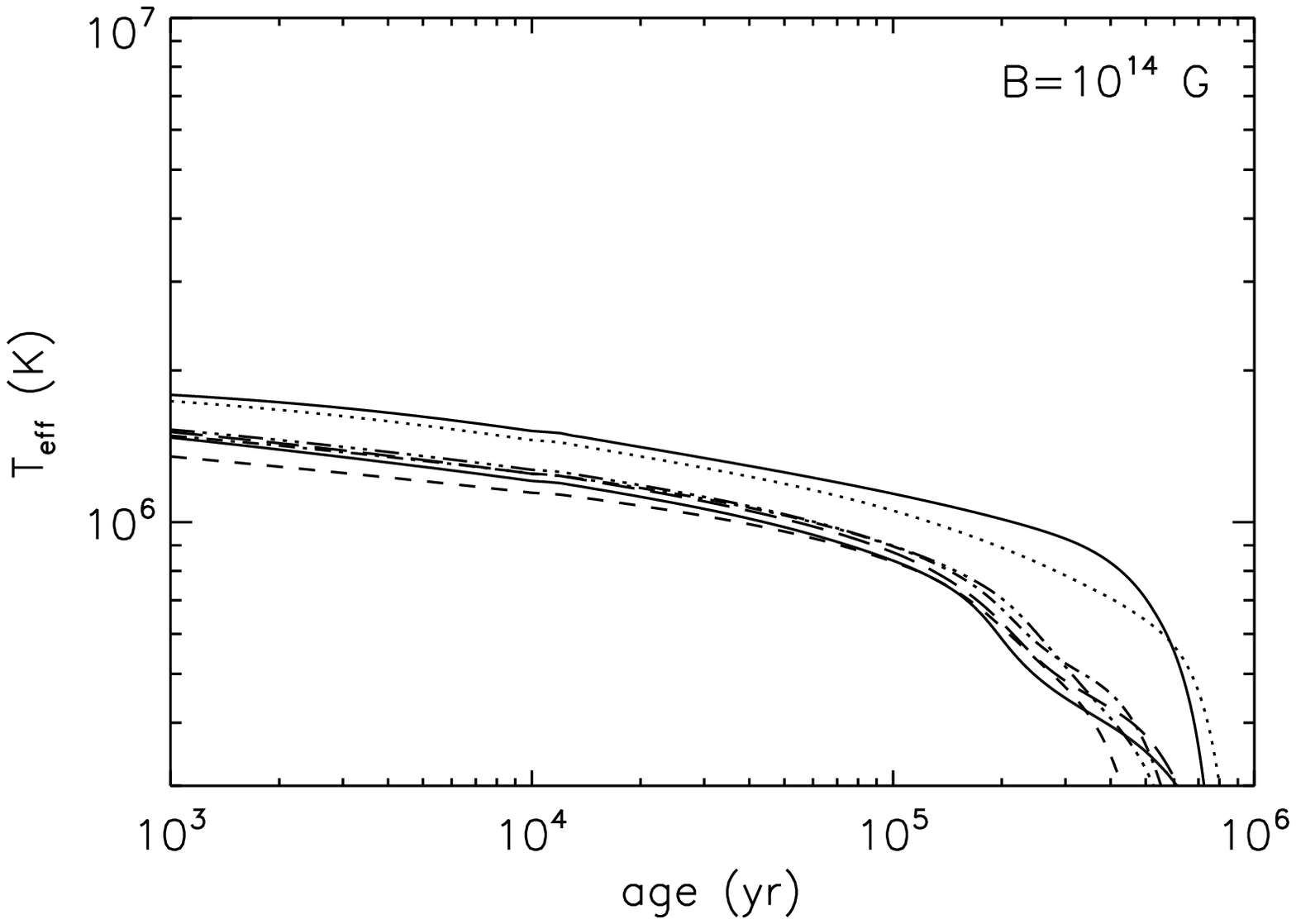} 
\caption{Left: Comparison of cooling curves of an $M=1.25 M_\odot$ NS with different
values of the initial magnetic field strength. From bottom to top:
$B= 3 \times 10^{12}, 10^{13},  3 \times 10^{13}, 10^{14}, 3 \times 10^{14}, 10^{15},$ and $3 \times 10^{15}$ G.
Right: Comparison of cooling curves with $B=10^{14}$ G 
and different masses: $M=1.10$ (top solid line), 1.25 (dots), 1.32 (dashes), 
1.40 (dash-dot), 1.48 (dash-triple dot),
1.60 (long dashes), and 1.70 $M_\odot$ (bottom solid line).
}
\label{fig1}
\end{figure*}
%%%%%%%%%%%%%%%%%%%%%%%%%%%%%%%%%%%%%

In the left panel of Fig.~\ref{fig1}, we show a sample of cooling curves (effective temperature versus
true age) for an $M=1.25 M_\odot$ NS but varying the initial strength 
of the magnetic field. For $B>10^{13}$ G the presence of strong magnetic fields
has a visible effect from the very beginning of the evolution (results for a  $M=1.4 M_\odot$
are very similar to this case).
The effective temperature of a young,
$t=10^3$ yr magnetar with $B>10^{15}$ G is a few times higher than that of a
NS with a standard $B=10^{13}$ G, and it is kept  above $10^6$ K for a much longer time. 
The effect is qualitatively similar, although somewhat smaller, for high mass stars, and of
course it also depends on details about the microphysics input. The influence of the 
superfluid neutron gap in the core is particularly relevant, in this work we use the results from
\cite{Baldo1998}. We also use an updated $^1S_0$ neutron superfluid gap in the crust
obtained from Quantum Monte Carlo simulations \citep{Carlson:2008}. 
This results in small
differences when comparing to the results in  \cite{PMG2009}.
In non-magnetized NSs, a smaller gap results in higher
temperatures at early times (suppression of neutrino emissivity),  but varying the gap 
does not change significantly the temperature of magnetars. The purpose of this paper is to study the observational
implications of different initial magnetic fields, being all the rest equal, so hereafter we
will fix the underlying physical model and we will only vary the normalization of the field
(not its geometry) and the mass of the NS. 
The variability of the cooling curves on the mass for a fixed magnetic field
is shown in the right panel of Fig. \ref{fig1}. We have chosen $B=10^{14}$ G and varied the
mass of the NS in a wide range, namely $M=1.10, 1,25, 1.32, 1.40, 1.48,
1.60, 1.70,$ and 1.76 $M_\odot$. Except for the two lowest masses, the rest
of cooling curves are difficult to be distinguished. This is clearly different from the
case of non-magnetized NSs, where
there is a clear separation in two scenarios: standard cooling (low mass) and
rapid cooling ($M>1.6 M_\odot$ in our model) discussed extensively by other
authors \citep{Yakovlev2004,Page2006}. In magnetized NS the fast cooling
scenario is masked by magnetic heating, becoming hard to distinguish whether
or not a fast neutrino emission process is active in high mass stars, as
pointed out in \cite{Aguilera2008b}. This raises an important
issue: disentangling the magnetic field initial distribution function is needed before we
can actually constrain other physical parameters that affect neutron star
cooling. 
For this purpose, we use a population synthesis technique 
that accounts for joint statistical properties of a set of objects, 
rather than trying  to fit individual objects with particular cooling models.

As in the case of the cooling scenario without extra heating by magnetic field decay, 
objects with the smallest masses are hotter in average.
In the mass distribution function we use they are very abundant and, since they are
more easily detectable because of their larger thermal luminosities (for the same
field range), most of the observed objects must be low mass NSs. 
 For NSs with $M>1.3M_\odot$ the effect of changing mass is
controversial, since it strongly depends on the assumptions about neutrino fast cooling processes.
Comparison between the left and right panels of Fig.\ref{fig1}  shows
that changes in magnetic field by a factor of $\sim 3$ are more important than
changing from $1.3M_\odot$ to $1.76 M_\odot$ for a fixed magnetic field. This implies that 
population synthesis studies are more sensitive to varying the initial magnetic field distribution than
varying the NS mass distribution, unless the population of low-field ($<10^{13}$~G) compact objects is discussed.
Normally, one would expect that isolated, very cool objects are low field and high mass NSs, so these would be
the best candidates to test fast cooling mechanisms without the complications due to the presence of strong magnetic fields.

%%%%%%%%%%%%%%%%%%%%%%%%%%%%%%%%%%%%%
\begin{figure}
\includegraphics[width=230pt,angle=0]{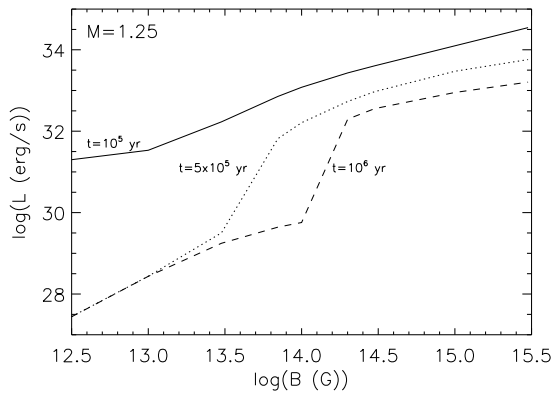} 
\caption{Luminosity as a function of the initial magnetic field strength for a $M=1.25 M_\odot$ NS at different
ages ($10^5$, $5\times10^5$ , and $10^6$ years).}
\label{fig2}
\end{figure}
%%%%%%%%%%%%%%%%%%%%%%%%%%%%%%%%%%%%%

 When we compare our model calculations with observations, 
we confront emission properties of NSs (observed vs. calculated fluxes or luminosities), not directly their fields,
but there is observational evidence for the correlation of these two magnitudes \citep{PonsLink2007}.
In the next section, when we establish constraints on the initial magnetic field distribution, the reader must keep in mind
that  these are model dependent constraints. What observations actually constrain
is the number of luminous objects, which we will translate to magnetic field strength using our theoretical models. 
As an example, in Fig. \ref{fig2} we plot the luminosity as a function of the magnetic field strength for a NS model
with $M=1.25 M_\odot$ and at different ages. While the luminosity of young ($<10^5$ yr) objects is less dependent on the 
magnetic field strength and relatively high (and therefore young NSs are more
easily observed), the luminosity of middle aged NSs depends strongly of the magnetic field strength. For a given NS model of a certain age, 
the luminosity increases sharply above a certain value of the initial magnetic field, thus making more magnetized objects
more likely to be observed. We remind again that the $x$-axis indicates the initial value of the magnetic field at birth, not the
corresponding value at a given age, which is always smaller as discussed above.

%%%%%%%%%%%%%%%%%%%%%%%%%%%%%%%%%%%%%
\section{Log N~--~Log S distribution for nearby cooling NSs}
%%%%%%%%%%%%%%%%%%%%%%%%%%%%%%%%%%%%%

 Previously we performed several calculations of the Log N -- Log S
distribution for NSs in the solar proximity for different sets of cooling
curves. Here we use the model of thermal evolution described above which
includes the magnetic field decay. The main motivation is related to the
fact that in the standard picture magnetic fields of magnetars decay, and
for some of close-by NSs (the M7) there are indications that their fields
are $\sim 10^{13.5}$~G, so additional heating due to field decay can be
important. 
Magnetic fields of some of the M7 sources are estimated by two methods:
spectroscopy and spin-down rate. The first is based on the unconfirmed hypothesis that wide depressions in their spectra 
are due to proton cyclotron lines \citep{Haberl2004}. The second method applies the usual magneto-dipole braking formula, but this is only
possible when a measure of the spin period derivative is available. Estimates due to both methods provide more or less consistent 
results within a factor of  few. In our model with magnetic field decay, the typical strengths of about few$\times 10^{13}$~G are 
reached within few$\times 10^5$~yrs for initial fields $\sim10^{14}$~G, consistent with age estimates for these NSs 
(see \cite{Page2009} and references therein). 
Temperatures and spin periods of the M7 sources within our model are also consistent with such ages (see Sec. 6).

To calculate the Log N~--~Log S distribution
(number of objects $N$ with a flux above $S$) of
close-by isolated cooling NSs we use the Monte Carlo code developed before
\citep{popov2003, popov2005,posselt2008} that builds a synthetic
population of nearby isolated NSs. A general description of the population
synthesis technique can be found in \cite{pp2007}.
The main ingredients of the present model are: the spatial distribution of
NS progenitors, the interstellar medium (ISM) density distribution needed to calculate the
observed flux, the NS mass distribution function, a set of cooling curves, discussed in Sec.~2, and the
initial magnetic field distribution. We now comment on each of this points before 
discussing our results.

Progenitors of NSs are distributed according to the model used by
\cite{posselt2008}. The contribution due to close-by OB-associations (the
Gould Belt) is crucial. As before we consider the region up to 3 kpc from
the Sun.
Inside 500 pc we take as the distribution of progenitors the
distribution of massive stars from HIPPARCOS data \citep{hip1997}. 
Outside this volume, most
of NSs (243 per Myr out of the total number of 270 per Myr born inside 3
kpc) originate in one of the OB associations. Others are distributed in the
exponential Galactic disc.
 
 In our model for Log N -- Log S calculations, the NS formation rate (in units of object per pc$^2$ per year) 
is different at different distances from the Sun due to the Gould Belt contribution and the non-uniform distribution of OB associations. 
Therefore, it is not straightforward to extrapolate the local rate to obtain the total Galactic rate of SNae.
Using data from \cite{Tammann1994} we estimate that the value we
use ($\sim 250$ NS inside 3 kpc per Myr) can be rescaled
to $\sim$1.2 NSs per 100 years for the whole Galaxy. Here the number
250 is obtained by subtracting the additional contribution due to the
Belt ($\sim 2/3$ of all nearby NSs are produced in 600 pc around the Sun). 
We stress that the value 1.2 per 100 years is only a rough estimate.
Sources beyond $\sim 1$ kpc from the Sun do not contribute
much to the Log N -- Log S distribution, and in this region the NS formation rate in our model is twice larger than inside 3 kpc. 

As we are mostly interested in the Log N -- Log S behavior at ROSAT count
rates $>0.01$~--~$0.1$~cts~s$^{-1}$, the sources at $<1$~kpc dominate (those
born in the associations forming the Gould Belt and in
other not very far away OB associations).
That is why the global Galactic distribution (Galactic arms, etc.)
of NS progenitors is not important for our study with
applications to ROSAT cooling NSs.

For the ISM distribution we used an analytical
description, which was demonstrated to be successful before
\citep{popov2003,posselt2008}. Since here we do not intend to produce an
accurate full-sky map of the distribution of sources, and for computational
limitations, we do not consider more detailed models for the ISM 3D
distribution.

For the mass distribution we use one of the variants, presented in
\cite{posselt2008}. 
The spectrum is derived using HIPPARCOS \citep{hip1997} data about close-by
massive stars and calculations by \cite{whw2002,hws2005}.
We use eight mass bins (1.1, 1.25, 1.32, 1.4, 1.48, 1.6,
1.7, 1.76 $M_\odot$). The first two bins contribute $\sim$ 30\% each. The last two
less than 1\%. According to this distribution 90\% of NSs are born with $M<1.45 M_\odot$. 
Such mass spectrum is in agreement with mass measurements of the secondary components in double-NS binaries, 
(see \cite{Stairs2008} and references therein). These NSs never accreted and they can be accepted as good 
sources for initial mass determination (unless some effects of binary evolution are crucial). 
In \cite{posselt2008} it was shown that realistic manipulations with the mass
spectrum do not influence  Log N~--~Log S distributions significantly.

We accurately
integrate spatial trajectories using the bi-modal Maxwellian kick velocity
distribution \citep{acc2002} and the potential traditionally used in papers
on isolated NSs starting with \cite{p1990}. Nevertheless,
the velocity distribution of NSs and the Galactic potential
are not important ingredients for the results shown in this section, 
as we deal with young sources which do not move significantly from their birth places, 
and their velocities are nearly constant during this time. 
We tested several velocity distributions including the double-side
exponential with the mean velocity 380 km s$^{-1}$ proposed by
\cite{fgkaspi2006}. Our results are not sensitive for variation between
different velocity distributions which were proposed during last years and
which successfully explain the radio pulsar observations.
Also, velocity by itself does not influence the observability of sources
under study (in contrast with, for example, studies of isolated accreting
NSs). 

We calculate ROSAT counts for the Log N -- Log S plots assuming 
that the local emission
(there is a surface anisotropy consistent with the magneto-thermal evolution models)
is purely blackbody, i.e. we neglect any non-thermal contribution and we do not
consider effects of composition or magnetic fields in the atmosphere.
This is partly justified by the fact that the M7
dominate the sample, and their non-thermal emission is negligible \citep{Haberl2007}. 
Atmospheric and magnetospheric models can change the spectral energy distribution 
(not the total luminosity) 
significantly, resulting in differences in the observed flux at the detectors.
However, we think that our sample is too small, and our knowledge about average
properties of NS atmospheres is not mature enough to be included in a
population synthesis scenario. We suspect that taking this effects into
account would not change the results significantly, but in future works, with
better knowledge of NSs properties and a larger sample it will be necessary to
include these effects explicitly. We must also mention that we considered only
heavy element envelopes in our model. The effect of an accreted H-He envelope
was discussed in \cite{Page1997} and it will be worth exploring in future works.

Sources are observable in soft X-rays while they are hot. In our previous studies we used
cooling tracks till an object cools down to 100 000K. In this paper we
put the limit at 300 000K due to computational reasons. However, we tested
that this modification does not change our results, as we confront our
results with observations of relatively bright sources.

\begin{figure}
\includegraphics[width=230pt,angle=0]{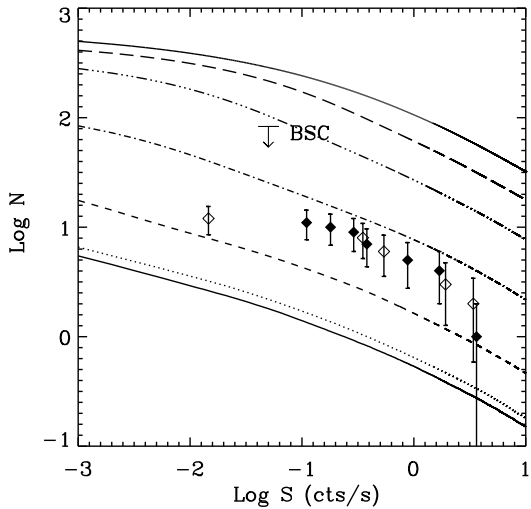} 
\caption{Log N~--~Log S distributions for different initial magnetic fields. From bottom to top:
$3\times 10^{12}$, $10^{13}$, $3\times 10^{13}$, $10^{14}$, 
$3\times 10^{14}$, $10^{15}$, $3\times 10^{15}$~G. 
``BSC'' -- is an upper limit from the ROSAT Bright Source Catalogue.
Filled symbols correspond to an addition to the distribution
one of the M7 sources, empty symbols correspond to an addition of PSRs 
(including ``the second Geminga'', see \citet{Mirabal2001, hcg2007}).}
\label{lnls}
\end{figure}

Results of the population synthesis modeling are applied to the ROSAT
all-sky survey, which is the most complete survey available at the energy
range of interest (0.1-1 keV) for our research, and the most uniform sample of close-by
young objects suitable to be used in population synthesis calculations. For
each NS ``observed'' in the simulation we calculate ROSAT PSPC counts per
second ($S$). Log N~--~Log S distribution for these sources is shown in
Figs.\ref{lnls}, \ref{lnls_aver}. 
Filled symbols correspond to additions of one of the M7 NSs,
empty symbols correspond to other sources -- Vela, Geminga, B1055-52,
B0656+14, and  3EG J1835+5918 (the second Geminga). 
Note, that for the last five sources we
plot total ROSAT counts (i.e. blackbody plus non-thermal magnetospheric
emission). If we include only the thermal component, then the shape of the
observed Log N~--~Log S distribution is slightly changed, but clearly the
first and the last filled points will remain as they are, and the change
in the shape is not significant. 
Also, non-thermal contributions to luminosities of these PSRs are not large \citep{bt1997}.
Error bars represent poissonian
statistical errors (square root of the number of sources). ``BSC'' is the
upper limit from the ROSAT Bright Source Catalogue \citep{voges1999}.
In the second Log N -- Log S plot we 
also show as a horizontal line the most recent limit 
at $90\%$ confidence level:  $N<31$ for soft, non-variable NSs, 
and $N<46$ for all NSs, including hard and variable sources 
(M. Turner et al., in preparation, R. Rutledge, private comm.).
We do not include magnetars in this sample. 
If we had considered further regions of the Galaxy
(birth place beyond 3 kpc from the Sun), 
we should have included the additional contribution due to young
magnetars, which can be visible from very large distances 
(all known young active magnetars are at 
distances $\ga 3$~kpc, according to the McGill SGR/AXP online catalogue 
\footnote{\rm http://www.physics.mcgill.ca/~pulsar/magnetar/main.html}).

%%%%%%%%%%%%%%%%%%%%%%%%%%%%%%%%%%%%%
%%%%%%%%%%%%%%%%%%%%%%%%%%%%%%%%%%%%%

In Fig.\ref{lnls} we show seven curves, each one calculated for a single
value of the magnetic field at birth (i.e. all NSs in the modeled population have
the same field). This illustrative graph demonstrates that, for our NS model,
low-field NSs ($B<3\times10^{13}$~G) cannot explain the observed
sources. If all NSs were born with the same initial field, it should be in
the range $3\times10^{13} - 10^{14}$~G. Larger initial fields result in
hotter NSs, and therefore in a large number of detectable sources. 
Here we focus on close-by NSs with fluxes above $\sim$0.1 ROSAT PSPC counts
per second, as in this range identification of NSs among ROSAT sources 
is considered to be mostly complete 
(see, for example, \citealt{sch1999,cag2002} and references therein). This
is also confirmed by  identification efforts like those by e.g.
\cite{a2006,c2005} and references therein.
%\citep{posselt2008}.
At lower fluxes ($<0.1$ cts~s$^{-1}$) many sources might be non-identified, yet. 
The ``second Geminga'' (the source with the smallest count rate) is $\gamma$-ray selected,
and this point must be taken as a lower limit for the Log N -- Log S
distribution at this flux range.
Clearly, multi-wavelength studies (cross correlation between catalogues obtained in different bands) are necessary
to find more close-by cooling NSs using the existing data, 
although one of the most crucial issues for identification is probably
the X-ray positional accuracy. Efforts in this direction, such as in \cite{Rut2008}, will be very useful.

%%%%%%%%%%%%%%%%%%%%%%%%%%%%%%%%%%%%%
\begin{figure}
\includegraphics[width=230pt,angle=0]{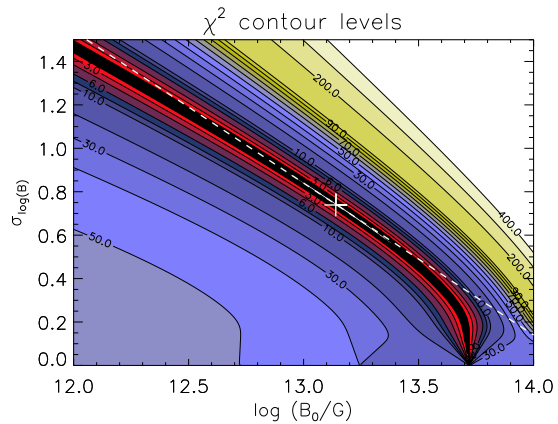} 
\caption{
$\chi^2$ (for 10 degrees of freedom) contour levels of the fits to the  Log N~--~Log S curves in the two-parameter space 
 $\log (B_0/{\rm [G])}$ and $\sigma_{\log B_0}$. The dashed line corresponds to  8\% of NSs born with
$\log (B_0/[G]) > 14.2$. }
\label{chi2}
\end{figure}
%%%%%%%%%%%%%%%%%%%%%%%%%%%%%%%%%%%%%

From results in Fig.\ref{lnls} it is clear that, unlike in previous works,
now we have also to worry about the initial magnetic field distribution
(B-distribution), because now our cooling curves depend not only on masses
of NSs but also on their fields, and the latter effect seems to be more
important.
For most of the paper, we have chosen the birth magnetic field to 
satisfy a log-normal distribution with central value $x_c \equiv \log B_0$ and standard
deviation $\sigma_{\log B_0}$ because this type of distribution reproduces the observed
distribution of PSRs.
The probability of a NS to be born with a magnetic field in the range
between $B_1$ and $B_2$ is then
\begin{equation}
\frac{1}{\sqrt{2 \pi} \sigma_{\log B}}
\int_{\log{B1}}^{\log{B_2}}  \exp{\left\{-\frac{(x-x_c)^2}{2 \sigma_{\log B}^2}\right\}} dx =
\end{equation}
$$
\frac{1}{2}\left[ {\rm erf}\left(\frac{\log{B_2}-x_c}{\sqrt{2} \sigma_{\log B}}\right) - {\rm erf}\left(\frac{\log{B_1}-x_c}{\sqrt{2} \sigma_{\log B}}\right) \right]
$$
where $x=\log{B}$ and ${\rm erf}(x)$ is the error function. 

%%%%%%%%%%%%%%%%%%%%%%%%%%%%%%%%%%%%%
\begin{table*}
\caption{Magnetic field distributions used in this work. Three of them (G1,G2,G3) are log-normal distributions
defined by its central value ($x_c$) and width ($\sigma_{\log B} $). Numbers in the table indicate the fraction of NSs
in a given B-field bin centered in ${\log B}=x_c$.  The other distributions (A1,A2, No mag) are discrete distributions and the numbers indicate
the fraction of NSs with that particular field strength. The right column defines the linestyle used for each distribution in the plots. }
\begin{center}
\begin{tabular}{lcccccccccc}
\hline
Model & $\sigma_{\log B} $& $x_c$& $3 \times 10^{12}$~G &  $10^{13}$~G & $3
\times 10^{13}$~G &  $10^{14}$~G &  $3 \times 10^{14}$~G &  $10^{15}$~G &  $3
\times 10^{15}$~G
& Line\\
\hline
 No mag   & &   & 0.5      & 0.5 & 0.0 & 0.0 & 0.0 & 0.0 & 0.0 &Long-dashed \\
A1    & &  &   0.3  & 0.2 & 0.1 & 0.1 & 0.1   & 0.1 & 0.1  & Solid\\
A2   & &   &    0.3 & 0.2 &0.2 & 0.1 & 0.1   & 0.1 & 0.0 & Dotted\\
G1  & 1.1 & 12.5 & 0.575 & 0.164&  0.114 & 0.08 & 0.039& 0.019& 0.009 &
Short-dashed\\
G2  & 0.84 & 13.0 & 0.37 & 0.244 & 0.191& 0.126 & 0.049& 0.0165& 0.0038 &
Dot-dashed\\
G3  & 0.46 & 13.5& 0.045& 0.243& 0.396& 0.263& 0.049& 0.0039&  0.000075 &
Dot-dot-dashed\\
\hline
\label{B-dist}
\end{tabular}
\end{center}
\end{table*}
%%%%%%%%%%%%%%%%%%%%%%%%%%%%%%%%%%%%%

In Fig. \ref{chi2} we show contour plots of the $\chi^2$ distribution of
fits to the Log N~--~Log S curves in the two-parameter space 
$\log B_0$ and $\sigma_{\log B_0}$, where $B_0$ is
given in Gauss.
The best fit to the observational data (cross) obtained with the IDL procedure CURVEFIT is
$\log (B_0/[G])=13.14$ and $\sigma_{\log B_0}=0.74$ but it is clear
that the results of the fit are highly degenerate, and any pair of
parameters located in the central diagonal band are allowed. In the figure,
we also show the line that corresponds to the family of log-normal distribution functions 
with an 8\% of NSs born with $\log (B_0/[G]) > 14.2$ ($B_0> 1.6 \times 10^{14}$~G), 
which interestingly is nearly parallel to the band of lowest $\chi^2$. 
 We should stress again that this is a model-dependent result. For our given NS evolution
model, this means that actually Log N~--~Log S curves are not
constraining independently the average field at birth or its dispersion 
(the birth parameters determine the whole evolution), but
simply the number of NSs born as high-luminosity objects, which according to our
underlying physical cooling model is correlated to the number of magnetars.
For other cooling models not drastically different from ours, any
reasonable field distribution with approximately 8-10\% of NS born as
magnetars should also be acceptable.  This is consistent with the fact the we do
not observe any nearby ($<$ 3 kpc) magnetar.  It is important to note that,
since magnetars are visible for a long time from large distances,
results are so sensitive to the addition of more magnetars that even with $\sim$10 observed 
sources we can place constraints on the fraction of magnetars.

In Fig.\ref{lnls_aver}
we present Log N~--~Log S curves for six B-distributions (see the Table).
Three log-normal distributions (G1, G2, G3), as they are selected from the
best fit for Log~N~--~Log~S, pass through the observed points. 
For comparison, we also use several other variants of the B-distribution, 
summarized in the Table. 
Values in the Table correspond to fractions in each magnetic field bin
normalized to unity.
The first one is an extreme case with no NS with fields above $10^{13}$~G. 
The other two are ``hand-made'' distributions (A1 and A2) 
in both of which 1/2 of NSs belong to the
PSR-range fields, and the rest is distributed among high-field objects.
The curve for model A2 (dotted line) shows that a 30\% of NS with
magnetic fields $\geq10^{14}$ G is already in contradiction with observations of the local population of cooling NSs. 
Addition of more NSs with very large initial magnetic fields (i.e. model A1) largely
overpredicts  the number of nearby objects detected as bright thermal sources. 

In the presented graphs all Log N~--~Log S distributions are plotted for
5000 calculated tracks, each of which is used for all eight masses, and all
data along a track is used with a time step $10^4$~yrs. So, the results are
significantly smoothed. We made a few additional runs for realistic numbers of
NSs (810 NSs born during 3 Myrs in the calculated region up to 3 kpc
from the Sun). 
Of course, such Log N -- Log S distributions are much more noisy.
However, statistical fluctuations cannot change our qualitative results.
For example, poissonian error bars for the data points typically bound curves
for G1, G2, and G3 distributions. Curves for magnetic field distributions
with a significant amount of magnetars cannot explain the data even taking
into account statistical fluctuations (unless, of course, a very rare strong
fluctuation happened).

%%%%%%%%%%%%%%%%%%%%%%%%%%%%%%%%%%%%%
\begin{figure}
\includegraphics[width=230pt,angle=0]{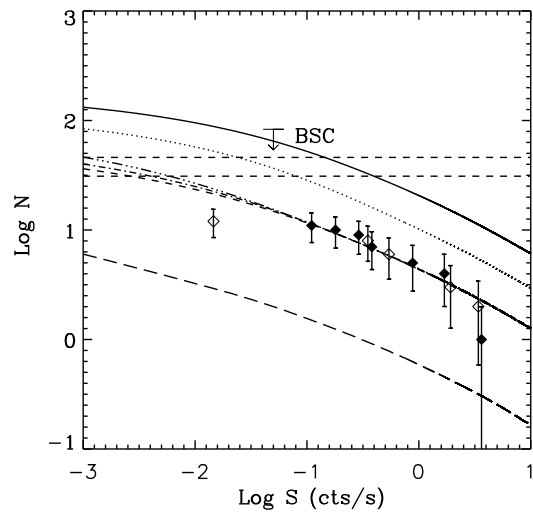} 
\caption{ Log N~--~Log S distributions for six variants of B-distributions. From top to bottom:
A1, A2, G3, G2, G1, No mag. See the Table for description of models and curve
styles. In this plot we also added a horizontal dashed lines which
correspond to 46 and 31 sources (see the text). This is an upper limit by M. Turner et al.
(in prep.; B. Rutledge private communication).} 
\label{lnls_aver}
\end{figure}
%%%%%%%%%%%%%%%%%%%%%%%%%%%%%%%%%%%%%

Having in mind that observational data can be fitted by different field
distributions, and that there are other important parameters not explored
(starting from superfluid gaps and ending with properties of the ISM), the
important message is that we can reproduce the data on the Log N~--~Log S
plot with realistic distributions, a normal NS model, and without fine
tuning. This gives us an opportunity to put a constraint on the number of
magnetars.  For our magneto-thermal evolution model,
initial field distributions with more than 30\% of NSs with initial fields above
$10^{14}$~G (even if all others have low fields) can be ruled out. 
 Assuming a log-normal initial field distribution, this also implies that
the fraction of NSs born with $B>10^{15}$~G can hardly be above a few percent.
Varying other parameters of the model can modify this conclusion, but not
dramatically, unless we allow for significant variations of the birth rate.

%%%%%%%%%%%%%%%%%%%%%%%%%%%%%%%%%%%%%
\section{Log N~--~Log L distribution for Galactic magnetars}
%%%%%%%%%%%%%%%%%%%%%%%%%%%%%%%%%%%%%

Motivated by a recent important paper by \cite{muno2008}, 
and as a consistency
check of the results of the previous section, we also consider a
simple model for the Log N~--~Log L distribution of highly magnetized NSs.
In \cite{muno2008} the authors analyze a large set of Chandra and XMM-Newton
data (nearly one thousand exposures) to search for new magnetars 
looking for pulsating sources in the range 5-200 seconds. No new
candidate was found, but this fact could be used to place upper limits on
the total number of Galactic magnetars with different properties. These
authors estimate that the number of magnetars with $L>3 \times
10^{33}$~erg~s$^{-1}$ and pulsed fraction larger than 15\% is below 540, and
the number of easily-detectable magnetars is $59^{+92}_{-32}$. Of course,
some (perhaps many) magnetars can have lower pulsed fractions and not being
detected as X-ray PSRs. In this respect the numbers by \cite{muno2008}
are representative of a fraction of the total population, and total limits
maybe larger than claimed. These are, to our knowledge, the best
observational limits on the number of such objects. On the other hand, the
fact that we observe some magnetars can be used to place lower limits. These
limits are too weak to favor one model against another, but it is
illustrative to compare some of the models that best fit the close-by,
intermediate field NSs to the magnetar population.

In our  model the  luminosity is of thermal origin, because we do not
consider magnetospheric processes that may cause the non-thermal emission.
However, since in magnetars both thermal and non-thermal components are supposed to be
powered by magnetic field decay, and energy conservation is satisfied by
construction in our evolutionary models, we can conclude that the total
energy release is calculated correctly, although the spectrum can be
different.

\begin{figure}
\includegraphics[width=230pt,angle=0]{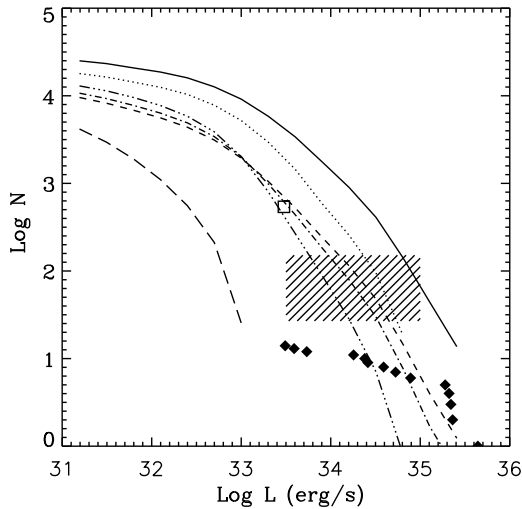} 
\caption{ Theoretical Log N~--~Log L curves for Galactic magnetars compared
with observational data and constraints. Diamonds show the Log N~--~Log L 
distribution for known magnetars (from the McGill group on-line
catalogue), the square indicates the limit of 540 weak AXP (Muno et
al. 2008), and the box corresponds to the estimated number of ''easily
detectable magnetars'' (Muno et al. 2008). We use the same linestyle for B-distribution models as
in Fig.\ref{lnls_aver} (see the Table).} 
\label{lnll_aver}
\end{figure}

We must stress that in this part no Monte Carlo simulation is done. Instead,
we use complete cooling tracks of NSs with different masses and magnetic
fields to estimate the whole Galactic population of NSs with a given
luminosity. Absolute numbers are obtained by normalization to the total
birth rate of NSs. We use the Galactic NS formation rate equal to 1/30
yrs$^{-1}$. This is close to the upper limit for NS formation rate \cite{keane2008}.
The uncertainty in the NS formation rate is a factor $\sim 2-3$ 
(see also \citet{keane2008}
and references therein) which may shift curves in Fig.\ref{lnll_aver}.
This NS formation rate is not directly related to the rate used for the Log N -- Log S calculations above.
In the case of close-by cooling NSs the rate of NS formation is determined by the properties
of the Gould Belt and close-by ($<3$ kpc) OB-associations. Here, in the case of magnetars,
we are interested only in the global galactic rate of NS formation (see a similar discussion
in \citealt{gh2007}). 

As we calculate distribution in luminosity, not in flux, we
do not take into account interstellar absorption. Calculations are done for the same cooling
curves, B-distributions, and mass distribution as used for Log N~--~Log S
calculations. 
Formally, very nearby sources with low fields can also
contribute to the observed Log N~--~Log L distribution, but in the range of
luminosities we are interested in ($>$ few $\times 10^{33}$~erg~s$^{-1}$) their
contribution is negligible. 

In Fig.\ref{lnll_aver} we show Log N~--~Log L distributions calculated for
the same six B-distributions described in the Table. We compare our curves
with the data by \cite{muno2008}: the upper limit to the number of magnetars
with $L>3 \times 10^{33}$~erg~s$^{-1}$ and pulsed fraction larger than 15\%
is 540 and the number of easily-detectable magnetars is $59^{+92}_{-32}$. Since
Muno et al. (2008) consider as easily detectable magnetars two types of
objects: bright magnetars with small pulsed fraction, and dim, but with very
large pulsed fraction, we show this as a rectangular region through which
satisfactory models should pass. As for real observations, at the moment we
know 5 SGRs (plus candidates), and 10 AXPs (plus transient objects and
candidates, see the McGill group on-line catalogue). 
These are also shown in the figure (diamonds) for comparison. 
In the on-line catalogue the luminosity is given for the range 2-10 keV.
Some magnetars also demonstrate significant hard X-ray emission \citep{Mereghetti2008}. 
It is not included in the plot, but in log-scale the shift is not crucial.
Note that this sample is not complete, so it must be taken as a lower limit for
the prediction of theoretical models. At the very bright tail it is possible that the
sample is close to complete, because there should be no more very bright
$L> 10^{35}$~erg~s$^{-1}$ magnetars.

Modeled Log N~--~Log L distributions start to flatten at $L\sim
10^{33}$~erg~s$^{-1}$. This value corresponds to the weakest known
magnetars. The results from the previous section show that all acceptable log-normal 
distributions for middle-aged thermally emitting NSs are similar in this luminosity range. 
They all predict about a thousand NSs above this
luminosity. Note that we assume that all sources are persistent and the
models considered in this work only include steady magnetic field Ohmic
decay. In addition to purely Ohmic decay, \citet{PonsGeppert2007} found that
the Hall drift may contribute noticeably to accelerating the dissipation of
magnetic fields in young NSs. This Hall phase lasts a few $10^3$--$10^4$
years and is characterized by an intense exchange of magnetic energy between
the poloidal and toroidal components of the field and by the redistribution
of magnetic field energy between different scales. It can be expected that
such rearrangements and the relatively rapid field decay may enhance the
average luminosity and result in crustal breaking and active stages (bursts,
flares), as can be observed in magnetars. This would increase the number of
bright sources with $L \sim10^{35}$~erg~s$^{-1}$, and this expectation is
 preliminary confirmed by
some artificial models in which we tried to take into account the possibility of this
transient behavior in young highly magnetized NSs. Alternatively, one can
also consider a fraction of NSs with larger internal toroidal fields that
have larger luminosities, but this introduces yet another free parameter in
the problem. A careful study of the first few thousand years of an
ultra-luminous magnetar's life and its transient epochs is out of the scope
of this paper, that focuses on long-term evolution and statistics.

%%%%%%%%%%%%%%%%%%%%%%%%%%%%%%%%%%%%%
%%%%%%%%%%%%%%%%%%%%%%%%%%%%%%%%%%%%% 

%%%%%%%%%%%%%%%%%%%%%%%%%%%%%%%%%%%%%
\section{Evolution of PSRs and the $P~-~\dot P$ diagram}
%%%%%%%%%%%%%%%%%%%%%%%%%%%%%%%%%%%%%

We turn now to PSRs. We have performed Monte Carlo simulations to
generate a synthetic PSR population and confront our models with
observations. The methodology employed in the simulations closely follows
the work by \citet{fgkaspi2006} but some parameters of their model are
allowed to change according to the results of our previous sections. The
main goal of this section is to answer the following question: can we obtain
a synthetic PSR population compatible with the observed one and
consistent with our previous description for magnetars and close-by isolated
neutron stars ? To this end, we start from the optimal population model
parameters obtained by \citet{fgkaspi2006} and modify only the initial
period and magnetic field distributions to account for the effect of
magnetic field decay consistent with our model.

To generate the PSR synthetic population we first choose the parameters of
the NS at birth closely following the model described by \citet{fgkaspi2006}, which we
briefly summarize.
The age of the NS is chosen randomly in the
interval $[0,t_\mathrm{max}]$, where $t_\mathrm{max}=$ 500 Myr. This is shorter than the
age of the Galactic disk, but it is enough for our purpose because PSRs
older than this age have crossed the death line (assuming standard magnetic dipole
braking) and are no longer visible as
PSRs. The place of birth is obtained according to the distribution
of their progenitors (massive Population I stars) which are mainly
populating the Galactic disk and more precisely its arms. The velocity at birth is distributed
according the exponential distribution with a mean value of 380 km~s$^{-1}$.
\footnote{From the point of view
of velocity and initial spatial distributions these assumptions differ from
those used for Log N -- Log S calculations. However, as we describe in Sec.
3, Log N -- Log S calculations are not very sensitive to the velocity
distribution and to large scale spatial distribution. The first is due to
relatively small ages of studied sources. The second, because in Log N -- Log
S at significant ROSAT count rates close-by sources dominate. 
The same can be said about differences in the models for the Galactic
potential in $P$~--$\dot P$ and Log N~--~Log S calculations.}

The spin period of the star at
birth, $P_0$, is chosen from a normal distribution with a mean value of
$<P_0>$ and standard deviation
$\sigma_{P_0}$. Of course, only positive values are allowed. The initial
magnetic field at the magnetic pole is obtained from a log-normal distribution
with mean value $<\log (B_0/{\rm [G])}>$ and standard deviation $\sigma_{\log B_0}$ .

Once we have chosen the properties of the NS at birth we solve the
appropriate differential equations to obtain the position, period and
magnetic field at the present time. We use a smooth model for the Galactic gravitational
potential \citep{KG89,CI87}. The period evolution is obtained by assuming
that the rotation energy losses are due to magnetic dipolar emission (orthogonal rotators),
where the magnetic field is obtained from
our magneto-thermal evolutionary models described in section 2.

\begin{figure}
\includegraphics[width=230pt,angle=0]{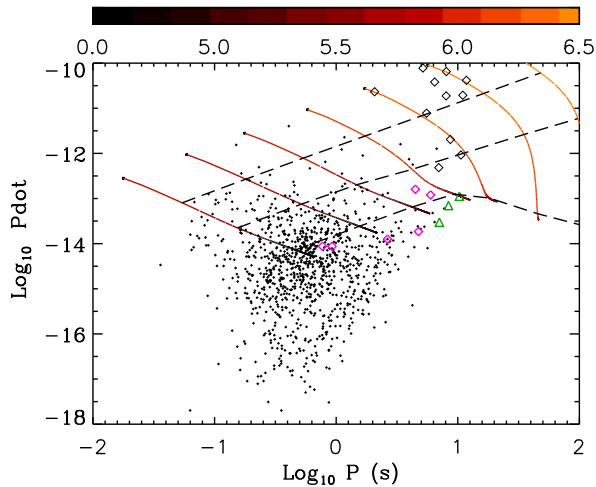}
\caption{$P-\dot{P}$ diagram for the sample of 977 real pulsars retained in our analysis
(small filled circles) with seven evolutionary tracks with different initial 
magnetic fields. For comparison, we also show AXPs and SGRs (open diamonds)
three of the M7 (triangles), and RRATs (filled diamonds). For RRATs new data from 
McLaughlin et al. (2009) is used. Color of the tracks
reflects the NS temperature, and long 
dashed lines indicate true ages of $10^4$, $10^5$ and $10^6$ yr. 
The color bar shows temperature in logarithmic scale.}
\label{ppdot}
\end{figure}

At the end of the Monte Carlo simulation we end up with a synthetic
population of PSRs to be compared with a given observed sample. We use the
PSRs detected in the Parkes Multibeam Survey (PMBS) sample
\citep{Lyne2008} and, to limit the
contamination of our sample by recycled PSRs, we further ignore the
PSRs with $P < 30$ ms or $\dot{P} < 0$, and those in binary systems. With
this restrictions, our resulting sample contains 977 objects. 
We use the parameters for detectability in the survey, radio luminosity and beaming
given in \citet{fgkaspi2006}. For comparison, The $P-\dot{P}$ diagram for the sample of 
real pulsars retained in our analysis is shown in Fig. \ref{ppdot}, together 
with some evolutionary tracks of the model used in the analysis. For lower fields,
NSs move nearly along constant magnetic field lines. We must remark again
that, in magnetars,  a somewhat enhanced field decay is expected to happen during the
initial Hall stage. This non-linear term is not yet included in our numerical simulations
and we expect a more vertical initial trajectory for objects in the upper right corner of the diagram.

%%%%%%%%%%%%%%%%%%%%%%%%%%%%%%%%%%%%%
%%%%%%%%%%%%%%%%%%%%%%%%%%%%%%%%%%%%%

The problem of defining an {\it optimal} model has been well discussed in
Sect. 3.7 of \citet{fgkaspi2006}, and we have adopted their approximate
approach that requires of some human judgment, rather than attempting to
cover a huge parameter space with our limited computational resources. We
have explored the parameter space in the region that best fits the
properties of thermally emitting NSs and magnetars, as described in previous
sections. Without pretending to perform a rigorous fully quantitative
analysis, we have considered the Kolmogorov-Smirnov goodness-of-fit test to
quantify our statistical analysis. 

\begin{figure*}
\includegraphics[width=230pt,angle=0]{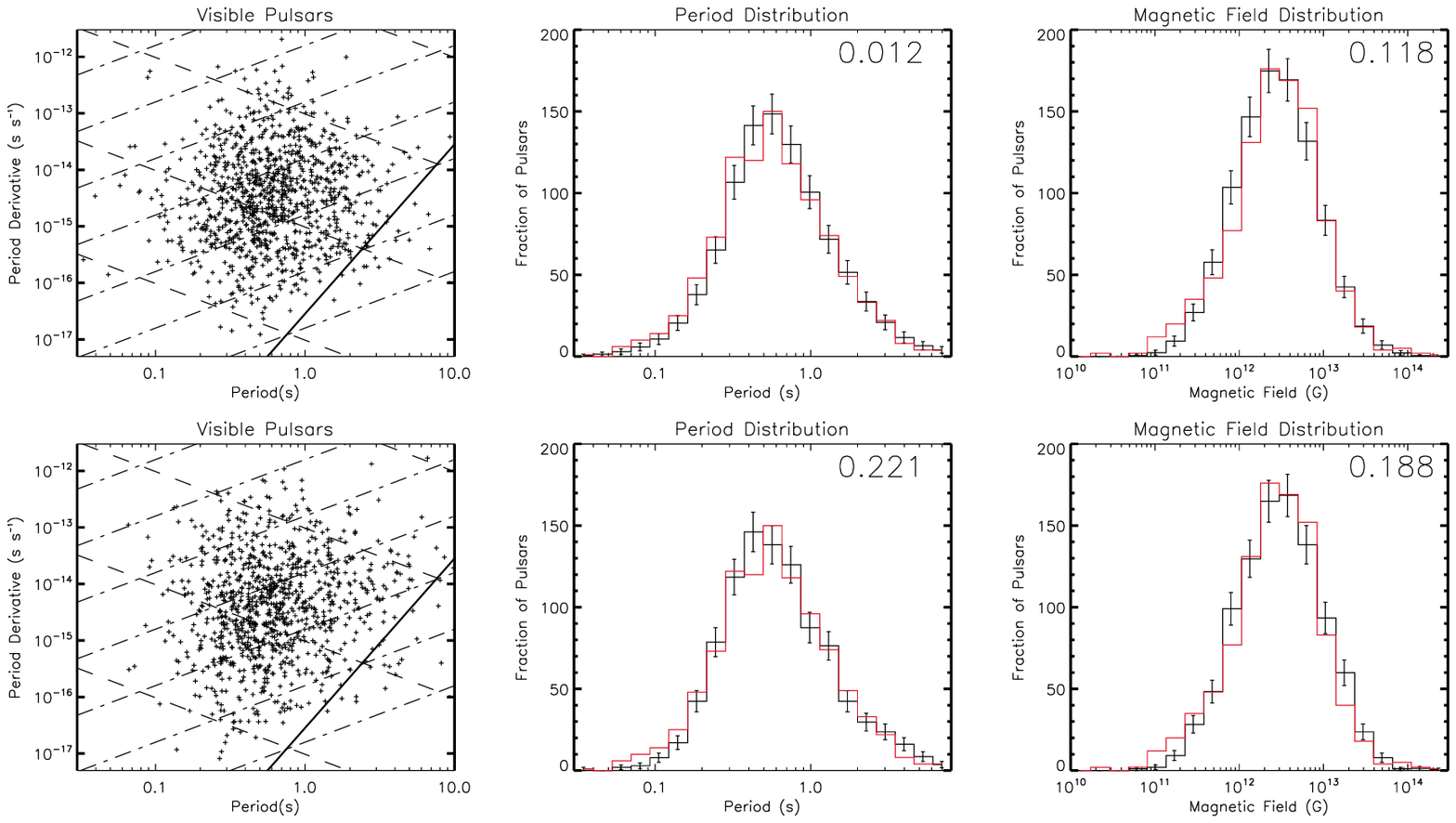}
\caption{ $P-\dot{P}$ diagram for  typical Monte Carlo realizations and distributions of observed PSR periods and
magnetic fields.  The distributions show the average of 50 realizations (error bars indicate standard deviation) compared to the PMBS sample
distribution (red lines).  The upper panels show the results for the optimal model without field decay \citep{fgkaspi2006},
and the lower panels correspond to the optimal model consistent with our simulations of magneto-thermal evolution
of NSs with field decay. On each histogram the associated Kolmogorov--Smirnov P--value is displayed in the upper right
corner. } 
\label{radiopsr}
\end{figure*}

In Fig. \ref{radiopsr} we show $P-\dot{P}$ diagrams for typical Monte Carlo
realizations and their corresponding distributions of observed PSR
periods and magnetic fields (averaged over 50 realizations of 977 detectable PSRs).
The upper and lower panels show the optimal models of \citet{fgkaspi2006}
and this work, respectively. The values for the distribution of the optimal
model assuming there is no field decay, are $<\log (B_0/{\rm [G]})>=12.95$ and
$\sigma_{\log B_0}$=0.55, $<P_0>=0.3$ s, and $\sigma_{P_0}=0.15$ s.
Note that \citet{fgkaspi2006} use the magnetic field at
the equator and the value they give is $<\log (B_0/{\rm [G])}>=12.65$.

Among all the models analyzed, we find
that our {\it optimal} model with realistic magneto-thermal evolution of NSs
corresponds to $<\log (B_0/{\rm G})>=13.25$,
$\sigma_{\log B_0}$=0.6, $<P_0>=0.25$ s, and $\sigma_{P_0}=0.1$ s. Since our
model includes field decay, we find that our average initial magnetic field
is about a factor of 2 larger than that of \citet{fgkaspi2006}, and the
distribution is also slightly wider. Because of this larger average field,
the initial period distribution has to be shifted to lower values to obtain
a visible synthetic PSR population statistically similar to the
observed population of PSRs. 

Our main conclusion is that, within
the parameter space that best fits the observed population of nearby,
thermally emitting NSs and magnetars, we can also find an optimal
parameterization that satisfactorily explains the observed PSR population. 
Therefore, it is possible to describe simultaneously
different families of the neutron star zoo with a single underlying physical
model. Interestingly, a combined statistical analysis of PSRs and
thermally emitting NSs allows to break the degeneracy in the parameter space that
arises when we try to work with a single family. Models with $<\log (B_0/{\rm
[G]})>=13.3$
or $<\log (B_0/{\rm [G]})>=13.2$, or with wider or narrower distributions
($\sigma_{\log B_0}$=0.7 or $\sigma_{\log B_0}$=0.5) give a much worse result
in the Kolmogorov-Smirnov test, obtaining in all these cases P-values $<0.01$ s.
Therefore, having fixed a NS model and initial magnetic field geometry,
and within the confidence region obtained in Log N -- Log S analysis, we conclude that
the observed distribution of radio PSRs is only consistent with values in a narrow vicinity of the optimal
model. In the future, as we improve the NS evolutionary models results may change, but the interesting
fact is that a combined analysis turns out to be very restrictive and breaks the degeneracy obtained in the
study of populations of only nearby thermally emitting NSs or Galactic magnetars.
We leave a more extensive study of the influence of these or other parameters (velocity
distribution, birth rates, etc.) for future work.

%%%%%%%%%%%%%%%%%%%%%%%%%%%%%%%%%%%%%
%%%%%%%%%%%%%%%%%%%%%%%%%%%%%%%%%%%%% 

%%%%%%%%%%%%%%%%%%%%%%%%%%%%%%%%%%%%% 
\section{Discussion and final remarks.}
%%%%%%%%%%%%%%%%%%%%%%%%%%%%%%%%%%%%% 

In this paper we presented a multi-component population synthesis study.
The final goal in this approach would be to make a complete population synthesis
with a unique NS physical model that consistently explains all known types of young NSs just varying
parameters such as the NS mass, age, or the strength and geometry of the magnetic field. 
Still, even after more sophisticated theoretical calculations are available, comparison with the data
will proceed by steps, confronting each piece of simulated data with observational data of some population,
in a similar way to this paper.
One reason for engaging multi-population studies can be illustrated as follows.
From Fig. 3 it is visible that the Log N -- Log S distribution can be explained by a
single field model (with moderate additional heating, in our framework).
However, the Log N -- Log L distribution for magnetars, as well as the $P$-$\dot P$ plot 
for radio pulsars, of course, cannot be explained by this model
for {\it different} reasons. In the first case case, because no bright magnetars will be observed;
in the latter, because it will be impossible to reproduce the main part of the pulsar population.

To summarize, we believe that the approach we use is well motivated by a necessity 
to have a natural model without {\it ad hoc} assumptions about different subpopulations.
Note that here we tested a very  ``smooth'' model, in a sense that we do not put by hand
initially distinct populations (magnetars, M7, PSRs, etc.).
In each part of our study (Log N -- Log S for the M7,  Log N -- Log L for magnetars, $P$--$\dot P$ for PSRs) 
we model ``just NSs'' using the same initial magnetic field distribution (which in our model 
is the main parameter), without specifying unique particular features for a given subpopulation,
as it  is usually done (see e.g.  \citealt{ptp2006,gh2007,keane2008}).

For example, XDINSs are not a separately defined class with initially distinct properties, but they just
appear as a population coming out from a smooth unique initial distribution, and with the following features:
\begin{itemize}
\item The magnetic field in these sources has significantly decayed from larger initial values, so no
magnetar-like activity is present;
\item Due to their large initial fields spin periods are long, so no
radio pulsar activity is observed (may be due to narrow beams);
\item They are still young enough  for the magnetic field to be still decaying
 ($<10^6$ yrs, according to \cite{PMG2009}) and heating the
crust, so that they can be observed as relatively bright thermal sources.
\end{itemize}

In our model a typical M7-like source has an initial field of $B \sim 10^{14}$~G. 
The period is $P\sim 7$~sec at a true age of $\sim 5 \, 10^5$~yrs (spin-down age $\sim (8-9) \, 10^5$~yrs). 
At this time such an object has $B\sim 4 \, 10^{13}$~G and $T\sim (6-7) \, 10^5$~K. 
The additional energy being input by field decay in these sources can be estimated in each case by multiplying
the average magnetic energy density ($\sim  10^{26}$ erg~cm$^{-3}$), by the crust volume ($ \sim 10^{18}$ cm$^3$)
and dividing by the typical Ohmic decay timescale under such conditions (crust density and temperature), which
is  $\sim 10^{13}$ s. This gives roughly $10^{30-31}$ erg~s$^{-1}$ available from magnetic field decay. A fraction
of this energy can be radiated in the form of neutrinos, but the energy reservoir to keep the star warm is still very important.

We tested average ages and typical parameters (periods, magnetic fields, etc.) of NSs which contribute to the range of Log N -- Log S between  0.1
and 10 cts~s$^{-1}$, where all observed sources are located. On average modeled NSs with inital fields 10$^{14}$--3 10$^{14}$~G which contribute a lot to this range have ages 2 10$^5$ -- 5 10$^5$ yrs and fields (at the moment of observation) $\sim 7\, 10^{13}$~G. Periods are distributed between $\sim$2-20 sec, in correspondence with observed properties of the M7. Of course, some fraction of lower field NSs are also found in this range, in correspondence with observations of cooling radio pulsars (Vela, Geminga etc.). Detailed study of the modeled population in the range 0.1-10 cts~s$^{-1}$ is in progress, and will be presented elsewhere.

In any population synthesis approach inevitably it is necessary to make
simplifications and to neglect some details. Partly, simplifications made in
this study are justified by low statistics of known sources, partly by uncertain properties
of objects under study.
One of the main problems in a population synthesis study
is related to possible correlations between different parameters. Some correlations are irrelevant
for our purposes (i.e., the correlation between direction of velocity and spin axis) but others can be crucial. 
For example, in our study we assume that masses,
initial magnetic fields and spin periods are not correlated. However, this
is very uncertain for all subclasses of NSs.  If these three parameters are
correlated our results may change. 
For magnetars it was proposed that they can have massive progenitors
\citep{muno2006}, and normally more massive progenitors are expected to
produce massive NSs \citep{whw2002}. 
Thus, we can expect a correlation between field and
mass for magnetars but not for other NS classes.
Such correlations, valid only for a not well identified part
of a population, are very hard to confirm or to rule out. In 
Fig. \ref{fig1} one can see that an increase in the value of
the magnetic field is much more influential (on the thermal evolution) than
changes in mass. That is why we think that moderate mass-field correlation
in the case of magnetars has very little influence on our results. Other effects like
rotation and mass-loss would complicate the situation even more.

We also neglect all possible effects related to the fact
that a significant fraction of NSs are born in binary systems. Potentially, this
can be important for magnetars if their large magnetic fields are generated by amplification due to
rapid rotation of the protoNS. The precise mechanism of magnetar formation is still unknown.
In some models \citep{pp2006,bp2009} magnetars are born only in binary
systems where a progenitor core was spun-up due to accretion or tidal
interaction, similar to the main scenario for gamma-ray burst progenitors.

In our simplified standard scenario (it is clearly visible in Fig.\ref{ppdot})
the M7 sources do not seem to be descendants of extreme magnetars, but the
evolutionary link with some of the AXPs is possible. 
However, due to numerical limitations and the very different timescales,
our model does not include the possibility of enhanced magnetic field
dissipation during the fast initial Hall stage, 
nor transient phenomena that change
a quiet X-ray emitter into an active bursting source, and back. Understanding the short term violent behavior of young magnetars
may help to reconcile even the highest field objects with the M7. 

One possibility to test this hypothesis is to
look at the velocity distribution of these types of sources. Indeed,
velocity is a good invariant on the time scale of $\sim$1 Myr. If AXPs are
rapidly moving objects, as it was popular to assume some years ago, but M7
have relatively small velocities, then one would conclude that the two
populations are not related.  Interestingly, there is a recent measure of the
transverse velocity of the magnetar XTE J1810-197 \citep{v1810}.
The measured velocity is slightly below the average for normal young neutron stars,
indicating that the mechanism of magnetar birth need not lead to high NS velocities. 
 On the other hand, recently \cite{Motch2009} reported new velocity measurements for the M7 sources, 
and one of them (RX J1308.6+2127) appears to be a fast object.
 In the near future we expect that the
measurement of proper motions of some of these sources with new observations 
will contribute to our understanding of their evolutionary link, if any.

In our study we also assumed that the magnetic field structure is similar for all NSs
and we only rescale the normalization value for a fixed geometry. This is obviously
an arbitrary choice. There are several issues related with the magnetic field geometry
that need  further investigation. If the magnetic field is not supported by currents located in the crust, but 
instead by superconducting currents in the core, the field will dissipate on much longer
timescales, and the heating mechanism we assume is less important. The amount
of energy stored in an internal toroidal field, compared to the corresponding energy of 
the measured dipolar component is also unknown, and it is an important parameter that can significantly alter the results.
We have chosen toroidal fields which maximum amplitude is a factor of 2 the dipole
estimate (however, the energy stored on the toroidal field is only about a 10\% because it
is confined to a small volume),  in agreement with recent studies on MHD equilibrium configurations
\citep{mhd1,mhd2}, but this remains an open question. Some models we tried with
toroidal fields one order of magnitude larger produced much hotter objects, and overpredicted
the observed number of isolated NSs and magnetars, unless we reduce significantly the
birth rate.

In this paper we have not attempted to adjust our cooling curves for low 
initial magnetic fields in such a way that the local population of normal
PSRs with detected thermal emission is perfectly explained. 
Future more extensive calculations exploring other parameters (superfluid gaps, crust physical properties, etc.)
can help to fit better the population of thermally emitting local NSs,
 and a better knowledge of the mass spectrum can be important for low-field stars, too.
For now, we can work with the known 7 XDINS (for which field decay
is probably important) and 4-5 normal close-by PSRs (i.e. lower field sources without additional heating).
In fact, the lowest curve in Fig.\ref{lnls_aver} predicts only 1-2 objects
with flux larger than 0.1 cts~s$^{-1}$, while we know four near-by PSRs (Vela, Geminga, B1055--52, and
B0656+14). The low field cooling curves are more sensitive to details of the equation of state,
superfluid gaps, and neutrino emissivities (and the NS mass) than high field models, and changes
of this parameters (i.e. the neutron gap in the core) can shift up or down the curves.  Ideally, 
the low field population should be used in a separate analysis to constrain parameters of
the interior physics but, given the problem of very low statistics we are facing, it seems
meaningless to split our populations in more subgroups at this stage.
We made some tests to see if it can be done but, with the low statistics we manage and with the present day
uncertainties about details of NS cooling, we do not think that an extensive investigation of microphysical
parameters can contribute much to our understanding of NS properties. We certainly need more data before
making strong cases in favor of particular cooling models.

Despite many attempts (see \citealt{a2006,c2005} and
references therein) the number of M7-like sources is not increasing
significantly. More M7-like NSs can be found using deep XMM-Newton and
Chandra observations. A good example is the new source found by
\cite{pires2009}. Another possibility is to have a sample of
$\gamma$-ray sources selected  by Fermi/GLAST, but they should be not M7-like,
but PSRs with radio beams not pointing towards us (similar to the ``second Geminga''). The most promising 
way to  increase the number of M7-like sources is related to the
future eROSITA instrument aboard Russian satellite Spectrum-X-Gamma. New
sources are expected to be dimmer, younger, hotter and further away than
the seven ROSAT sources \citep{posselt2008}, and eROSITA (with non-truncated sensitivity at low
energies) will be a perfect tool for them. For magnetars there is some hope to
have many more candidates due to the MAXI instrument aboard the
International Space Station \citep{maxi2009}.
Then, with increased statistics, it will be useful to come back to study a separate Log N -- Log S distribution 
of close-by cooling PSRs.

To summarize, 
the methodology we employ in this article looks very promising to constrain NS properties as more data coming
from future missions arrive and more PSRs are found by radio surveys.
A future increase of the observational data will allow to perform
much more detailed population synthesis and to establish evolutionary links (if they exist)
between different classes of isolated NSs, and to
understand better the general evolution of these sources.

\section*{Acknowledgments} 
We thank D.N. Aguilera and R. Turolla for interesting discussions. We thank the referee for critical remarks. This
research has been supported by the Spanish MEC grant AYA 2007-67626-C03-02,
the Research Network Program {\it Compstar} funded by the ESF, and the
Russian Foundation for Basic research (grants 07-02-00961 and 09-02-00032). SP thanks
the Universities of Alicante and Padova for hospitality.

\bibliography{popov}

\end{document}